\documentclass{article}

\usepackage{arxiv}

\usepackage[utf8]{inputenc} 
\usepackage[T1]{fontenc}    
\usepackage{hyperref}       
\usepackage{url}            
\usepackage{booktabs}       
\usepackage{amsfonts}       
\usepackage{nicefrac}       
\usepackage{microtype}      
\usepackage{lipsum}
\usepackage{xspace}
\usepackage{graphicx}
\usepackage{amsmath}
\usepackage{chngcntr}

\title{Beyond the Coverage of Information Spreading: Analytical and Empirical Evidence of Re-exposure in Large-scale Online Social Networks}

\author{
  Xin Lu\footnotemark[1], \footnotemark[2]\\
  School of Business, Central South University, 410083 Changsha, China\\
  College of Systems Engineering, National University of Defense Technology, 410073 Changsha, China\\
  Department of Public Health Sciences, Karolinska Institutet, 17177 Stockholm, Sweden\\
  \texttt{xin.lu@flowminder.org} \\
   \And
 Shuo Qin,\footnotemark[2]\\
 College of Systems Engineering, National University of Defense Technology, 410073 Changsha, China\\
  \texttt{qinshuo17@nudt.edu.cn} \\
    \And
 Petter Holme \\
 Institute of Innovative Research, Tokyo Institute of Technology, 226-8503 Yokohama, Japan\\
 \texttt{pttrhlm@gmail.com} \\
    \And
 Fanhui Meng\\
 School of Data and Computer Science, Sun Yat-sen University, 510006 Guangzhou, China\\
 \texttt{mengfh3@mail2.sysu.edu.cn}\\
    \And
 Yanqing Hu \\
 School of Data and Computer Science, Sun Yat-sen University, 510006 Guangzhou, China\\
 Southern Marine Science and Engineering Guangdong Laboratory, Zhuhai, 519082, China\\
 \texttt{huyanq@mail.sysu.edu.cn}\\
    \And
 Fredrik Liljeros\\
 Department of Sociology, Stockholm University, 17177 Stockholm, Sweden\\
 \texttt{fredrik.liljeros@sociology.su.se}\\
    \And
 Gad Allon\footnotemark[1]\\
 The Wharton School, University of Pennsylvania, 19104 PA, USA\\
 \texttt{gadallon@wharton.upenn.edu}
}

\begin{document}
\maketitle

\begin{abstract}
Peer influence and social contagion are key denominators in the adoption and participation of information spreading, such as news propagation, word-of-mouth or viral marketing. In this study, we argue that it is biased to only focus on the scale and coverage of information spreading, and propose that the level of influence reinforcement, quantified by the re-exposure rate, i.e., the rate of individuals who are repeatedly exposed to the same information, should be considered together to measure the effectiveness of spreading. We show that local network structural characteristics significantly affects the probability of being exposed or re-exposed to the same information. After analyzing trending news on the super large-scale online network of Sina Weibo (China's Twitter) with 430 million connected users, we find a class of users with extremely low exposure rate, even they are following tens of thousands of others; and the re-exposure rate is substantially higher for news with more transmission waves and stronger secondary forwarding. While exposure and re-exposure rate typically grow together with the scale of spreading, we find exceptional cases where it is possible to achieve a high exposure rate while maintaining low re-exposure rate, or vice versa.
\end{abstract}


\section{Introduction}
Online social media is fundamental to the lives of billions of people all over the world, generating and sharing opinions, news on a massive scale. It meets the need for information acquisition, communication, entertainment through the integration of users' social circles in life and work, with the ability to create and share content in a frictionless manner. Studies of information spreading have focused so far on strategies for optimal coverage \cite{morone2015influence,aral2012identifying,kitsak2010identification,hu2018local,censor2010partial}, including selection of seeds, identification of the most efficient or influential spreaders, etc. Meanwhile, less attention has been given to the problem of information redundancy. Given the abundance of information, and the ease of sharing it, people are exposed repeatedly to the same information, arriving through different routes within their social networks. Recently, albeit, in a limited manner, the issue has been recognized as problematic. For example, a survey by Reuters of 1,300 corporate managers showed that information redundancy results in delays in making important decisions, and ultimately, results in lower job satisfaction\cite{edmunds2000problem}.

However, information redundancy may have positive effects. Improving re-coverage, and not only the coverage of a target audience is essential for the marketing of both products and ideas. With the increased resistance to traditional forms of advertising, marketers have turned to more organic forms of information propagation, and viral marketing through social networks with the holy grail of ``going viral'' or becoming a meme  \cite{iribarren2011branching}. Seeding strategies have been studied frequently in word-of-mouth (WOM) programs, and positive WOM is considered a powerful marketing medium for companies to influence consumers \cite{hinz2011seeding}. Studies have shown that mutual referrals between users can affect the user's preferences, thereby increasing a consumer propensity to purchase \cite{adamopoulos2015effectiveness,bapna2015your}. It was reported that 92\% of consumers believe suggestions from friends and family more than advertising, and beyond friends and family, 91\% of people trust online reviews written by other consumers as much as they trust recommendations from personal contacts \cite{nielsen2012consumer,local2014local}. When users are recommended by their friends one or more times, their psychological and behavioral outcomes may vary widely. Ref.~\cite{bakshy2012the} found that an individual is more likely to exhibit the sharing behavior when multiple friends share, even if she does not necessarily observe her friends' action. Ref.~\cite{centola2010the} find that individual adoption was much more likely when participants received social reinforcement from multiple neighbors in the social network when they investigate the effects of network structure by studying the spread of health behavior through artificially structured online communities. Other studies also show that social reinforcement played a crucial role in the online spreading of practices \cite{zheng2013spreading, granovetter1978threshold, Romero2011}.

We extend existing models and empirical efforts in the study of information spreading by adding a new dimension---the re-exposure rate. This measures the number of users who have been repeatedly exposed to the same information, and thus the redundancy of the information propagation process. We first approximate the redundancy probability of a node in the propagation process and show how the network structure plays a significant role in determining the information redundancy. Next, we investigate trending topics on a popular Chinese microblogging platform (Sina Weibo), with 430 million connected users, to validate factors on the likelihood of being exposed to the same message from multiple peers, including the number of followings and followers, the length of re-posting waves, and secondary re-posting, etc. Finally, we develop a simple but effective spreading strategy which could achieve near (minimum or maximum) re-exposure rate given a fixed exposure rate. From the managerial perspective, our findings provide constructive support for the control of redundant information on online social platforms. 

\begin{figure}
\centering
\includegraphics[width=16cm]{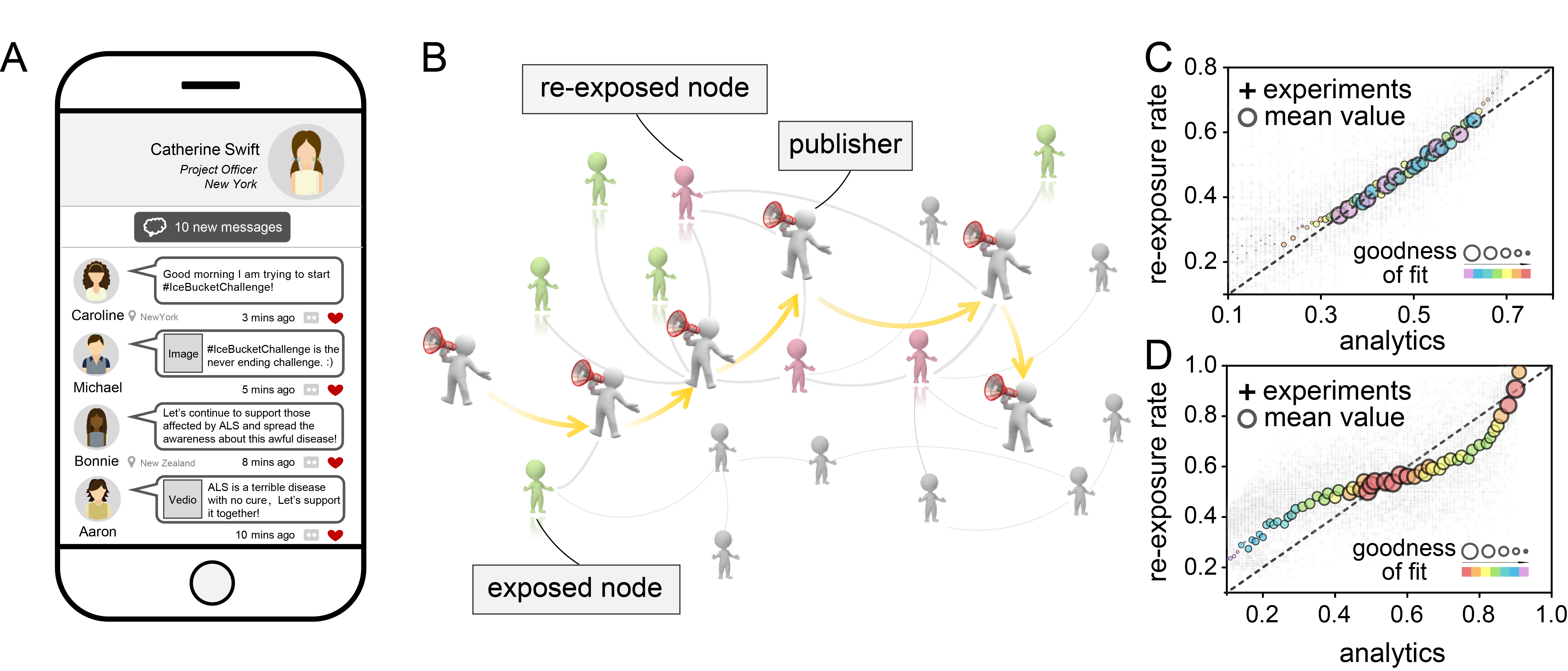}
\caption{($A$) An example of redundancy on social media. ($B$) A schematic chart for users who are exposed and users who are re-exposed by an information spreading process. ($C$) The agreement between the analytical value of node redundancy and the experimental value under different propagation models, including Random walk model and ($D$) SI model. The color and size of each circle is proportional to the goodness of fit (the distance between the mean value and the standard line).}
\label{fig:1}
\end{figure}

\section{Definition of exposure and re-exposure rates}
On most social media platforms, such as Facebook, Twitter, etc.\ \cite{zhang2016dynamics}, information redundancy is often described as repeated forwarding of news, posts or tweets, with a high frequency of repetition. For example, the ALS Bucket Challenge has gone viral in 2014, which was an online effort to raise awareness for people with Amyotrophic Lateral Sclerosis (ALS) disease and to raise funds. One of the reasons why this campaign became viral is that participants were asked to challenge minimum three persons to take the challenge, which create a viral loop. People shared their behaviors by posting videos, commenting and liking this topic, over 17 million people participating in the challenge worldwide and raise 11.4 million dollars in a span of just eight weeks. The craze has been spread quickly, resulting in users observing the same information when it was shared by different sources on their personal social networks, as illustrated in Fig.~\ref{fig:1}$A$. 

In the process of information spreading, the nodes can be divided into two categories (Fig.~\ref{fig:1}$B$). The ``exposed nodes'' are those with at least one of its friends posting the information; and the ``re-exposed nodes'' are those with two or more of their friends posting the same information. Let $G=(V,E)$ be the network with $V=|N|$ being the number of nodes and $E$ being the set of links, let $A={a_{ij}}$ be the adjacency matrix with $a_{ij}=1$ if $v_i$ and $v_j$ are connected, i.e., $i$ can see updates posted by $j$, and $a_{ij}=0$ otherwise. Let $\varepsilon(i)$ indicates the state of node, where $\varepsilon(i)=0$ indicates state S, which stands for the status that the node has not posted the information; and $\varepsilon(i)=1$ indicates state I, which represents that the node has posted the information, such that it can influence its friends of type S to become type I. We define:

\begin{equation}
\psi \left ( i \right )= \left\{\begin{array}{rcl}
 0, \sum\limits_{j=1}^{N} a_{ij}\varepsilon(j)< 1 &\\
 1, \sum\limits_{j=1}^{N} a_{ij}\varepsilon(j)\geq 1
\end{array}\right.
\end{equation}
to indicate whether individuals $i$ has any type I node among his social network; and use $\xi(i)$ to indicate whether $i$ is repeatedly exposed to the same information, i.e., has more than one type I friends:
\begin{equation}
\xi \left ( i \right )= \left\{\begin{array}{rcl}
 0,\sum\limits_{j=1}^{N} a_{ij}\varepsilon(j)< 2 &\\
 1,\sum\limits_{j=1}^{N} a_{ij}\varepsilon(j)\geq 2
\end{array}\right.
\end{equation}

Then the exposure rate ($R^{\rm exp}$) and re-exposure rate ($R^{\rm re}$) of the network is defined by the proportion of exposed nodes and re-exposed nodes, respectively.

\begin{equation}
R^{\rm exp}=\frac{\sum\limits_{i=1}^{N}\psi (i)}{N}
\end{equation}

\begin{equation}
R^{\rm re}=\frac{\sum\limits_{i=1}^{N}\xi (i)}{N}
\end{equation}

\section{Results}
\subsection{The analytics of exposure and re-exposure rates}
We use random walk \cite{pearson1905the} and branching process \cite{ney1965the} ({see {\textit {SI Appendix}}, section \uppercase\expandafter{\romannumeral1}}) to model the spreading of information on social networks and approximate $R^{\rm exp}$ and $R^{\rm re}$ based on the probability of visiting each node ({see {\textit {SI Appendix}}, section \uppercase\expandafter{\romannumeral2}}). Network data and experimental design are presented in {{\textit {SI Appendix}}, section \uppercase\expandafter{\romannumeral4}}. We can see that the theoretical value of $Pr_i$ (the probability that individual $i$ becomes a re-exposed node) agrees well with the experiments, most of the scatters fall near the standard line (Fig.~\ref{fig:1}$C$-$D$). Apart from this, as expected, we find that the degree of a node is positively correlated with $R^{\rm re}$, i.e., nodes with a large number of friends are more likely to get re-exposed to the spreading information. In the following sections, we also analyze the impact of other network structural  parameters on $R^{\rm exp}$ and $R^{\rm re}$.

\begin{figure}
\centering
\includegraphics[width=11.4cm]{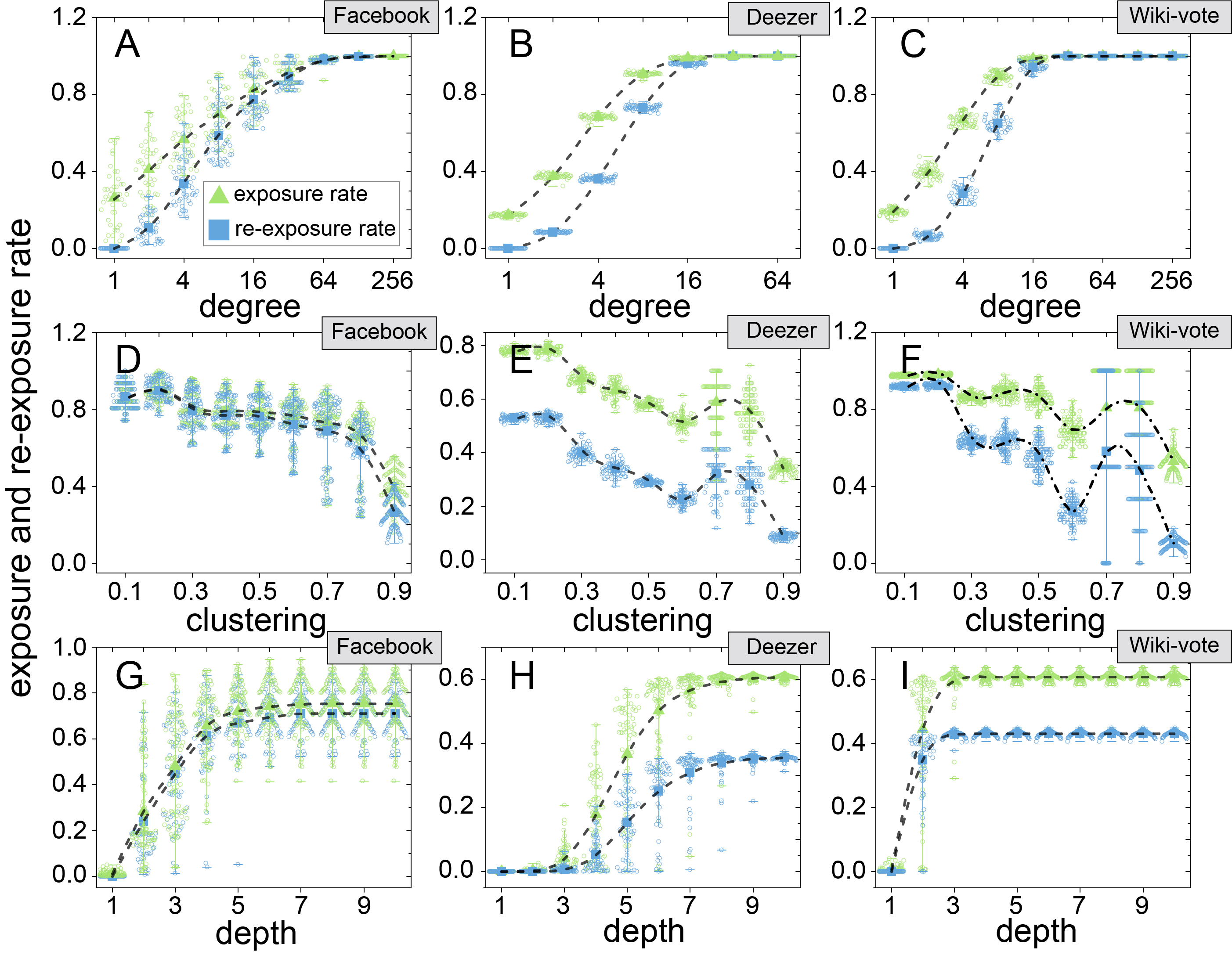}
\caption{($A$-$C$) The effect of degree , ($D$-$F$) clustering coefficient and  ($G$-$I$) depth on information exposure and re-exposure rate with 100 times simulated spreading on three empirical networks. The black line represents the mean value of each group of data.}
\label{fig:2}
\end{figure}

\subsection{Influence of network structure on information redundancy}
The correlation between node degrees and exposure rate ($R^{\rm exp}$), re-exposure rate ($R^{\rm re}$) when the information spreads to 20\% of the network nodes are presented in Fig.~\ref{fig:2}$A$-$C$ We can see that, in agreement with the analytical results, a larger degree is positively correlated with a higher likelihood of exposure and re-exposure rates. In all three empirical networks, nodes with a larger number of personal social connections are more likely to have access to new information (exposed) and are more likely to be repeatedly exposed.

While the degree of each node is essential as a driver of both exposure and re-exposure, it does not capture the depth of the relationship between nodes, and their likelihood information is going to be propagated on these links.  The ``strength of weak ties'' hypothesis argues that a network with many ``long ties'' spreads social behavior farther and more quickly than a network in which ties are highly clustered \cite{rogers1995diffusion,pennings1973measures,watts2000small}, and strong ties among family members and close friends are more likely to be activated for information flows and are also more influential than weak ties \cite{granovetter1973the}. In addition, communities play an important role in information diffusion and adoption, as users are associated with their like-minded users, redundant transmission pathways and additional recommendations from peers can substantially increase the likelihood of re-posting or re-sharing \cite{leskovec2007dynamics, centola2010the}. Therefore, clustering and community structures may create redundant transmission pathway, which in turn may limit the reach of information spreading. On the other hand, if a piece of information is repeatedly broadcast among the same group of interconnected users, we will see the emergence of an echo chamber, and a polarization behavior, as we indeed observe lately. 

The information spreading process can be characterized by the depth of re-post chains, with a re-post node of depth $w$ being $w-1$ waves away from the seed node \cite{liben2008tracing}. In Fig.~\ref{fig:2}$G$-$I$, we present the correlation between depths of re-post chains and exposure rate, re-exposure rate when the information spreads to 20\% of the network nodes. However, there is a notable difference for the relationship between $R^{\rm exp}$ and $R^{\rm re}$: in Facebook and Wiki-vote the $R^{\rm exp}$ and $R^{\rm re}$ grow faster than on Deezer. This can be attributed to the small average nodal degree in Deezer, which results in slower diffusion. Apart from this, while there is a constant gap (20\% to 30\%) between $R^{\rm exp}$ and $R^{\rm re}$, however this gap is not prominent on the Facebook network, which means clustered networks have more redundant ties, and high level of clustering will better promote the diffusion of behaviors.

\begin{figure}
\centering
\includegraphics[width=16cm]{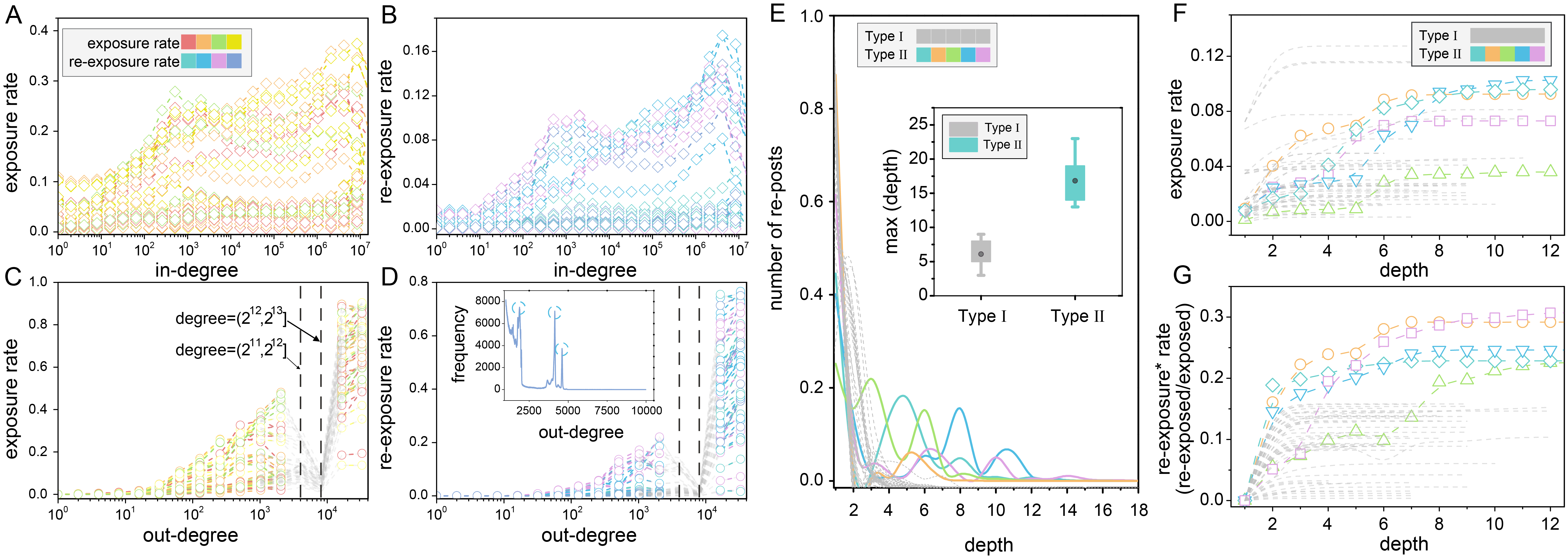}
\caption{ ($A$-$B$) Correlation between nodal in-degree, out-degree and exposure rate ,  ($C$-$D$) re-exposure rate of 50 trending topics in 2017 on Sina Weibo. Each curve represents a unique news item and the region between $2^{11}$ to $2^{13}$ are colored grey to highlight the lack of coverage and redundancy for commercial nodes and bot-supported-nodes in this region. ($E$) The number of re-posts , ($F$) exposure and ($G$) re-exposure rate change with the depth of re-post path for the 50 trending topics, respectively. The re-exposure$^*$ rate measures the ratio of re-exposed nodes out of all the exposed nodes. Grey and colored curves represent type I news and type II news, respectively.}
\label{fig:3}
\end{figure}

\subsection{Observation of information redundancy on a large-scale social network}
\subsubsection{Network structure and $R^{\rm exp}$, $R^{\rm re}$}
For the 50 most trending topics spread in Sina Weibo, we were able to map the comprehensive trajectory of each in the global network which contains 430 million active users and 51.8 billion links (see materials and methods). We then calculate and compare the nodal in-degree (Fig.~\ref{fig:3}$A$-$B$) and out-degree with exposure or re-exposure rate (Fig.~\ref{fig:3} $C$-$D$). We can see that in general both in-degree and out-degree are positively associated with $R^{\rm exp}$ and $R^{\rm re}$, for certain groups of nodes with higher in-degree and out degree, the highest $R^{\rm exp}$ can reach 37\% and 90\%, and $R^{\rm re}$ can be as high as 17\% and 78\%, respectively. There are however distinct difference between the pattern of in-degree and out-degree. While both $R^{\rm exp}$ and $R^{\rm re}$ do not increase significantly with the number of followers, they increase sharply with the number of people they are following.  

Surprisingly, we find a sudden drop in $R^{\rm exp}$ and $R^{\rm re}$ for nodes with out-degrees in the region of ($2^{11}$, $2^{13}$], indicating that nodes in this region have a substantially different local network structure compared to ``normal'' users. This phenomenon is mainly attributed to two reasons: first, users in this range may act with different purposes than an ordinary user. Following such a high number of users, which is above the two thousand people platform limited (above which a user has to apply for a special permit) may indicate a different motivation; second, these users may also be automated accounts or bots who randomly follow a large number of socially distant peers. We discuss this specifically in {{\textit {SI Appendix}}, section \uppercase\expandafter{\romannumeral5}}.

\subsubsection{Depth of spreading and $R^{\rm exp}$, $R^{\rm re}$}
The trajectory of transmission imposes constraints on the coverage of information spreading. Investigation the relationship between the depth of information diffusion and the number of re-posts reveals that the 50 popular news posts can be divided into two categories (Fig.~\ref{fig:3}$E$). For the first category (type I),  the number of re-posts declines with the depth of the information propagation trajectory; There are, however, noticeable fluctuations during the spreading process for the second category of news posts (type II). Another notable difference is that the maximum depth of type II news is 17, while it is only 6 for type I news. In a close examination on the content of these news, We find that the type I news are initially posted by influential users on the platform, e.g., pop stars, renowned scientists, famous writers, etc. These news posts are typically about advertising or personal promotion and the first wave of re-posting can accounts to 80\% of all sharing activity. Thus, the spreading of type I news is mainly due to the short relay of an influential micro-blogger's massive followers. On the other hand, the type II news are initially posted by less influential users, but they strike the right emotional chord with audience, contains valuable or attractive information, which means that the news may be re-posted by only a small number of initial followers, but it has the possibility of being forwarded deep into the network, and become viral when it has been re-posted by influential users.

The difference of the transmission pattern of type I and type II news is clearly shown in Fig.~\ref{fig:3}$F$-$G$. For a given level of coverage rate of $R^{\rm exp}$, there is no significant difference between type I and type II news, which has reached up to 13\% of the nodes in the network, i.e., 56.3 million users have seen the most populated news shared by their friends. However, among the exposed nodes, type II news have a substantially higher ratio of re-exposure rate, at the end of spreading, about 20\% - 30\% of nodes exposed to type II news have seen the same story more than one time. Interestingly, this ratio decreases to 0\% - 15\% for type I news. Also, we can observe that the growth of exposure rates and the modified re-exposure rates along the length of re-post waves are significantly different. While the $R^{\rm exp}$ and $R^{\rm re}$ of type I news increases smoothly until reaching a steady state with more waves of transmission, for type II news both $R^{\rm exp}$ and $R^{\rm re}$ are boosted by secondary sharings at particular stage, i.e., when some influential users have re-posted the news. This phenomenon can be illustrated by visualizing the transmission trajectories of two sample news (Fig.~S2). While the first is populated due to the massive number of immediate re-posts from the pop singer's followers, the second story spreads slowly at the beginning and only becomes viral when a super-connected user has forwarded it.

Based on our results, we can explain why some trending news are not known to most people, while others may flood our screens: the information with a strong tendency of being spread is more likely to enter the public view, rather than produced by paid posters (groups of internet ghostwriters paid to post online comments). This can be attributed to strong stimulation and reinforcements between users, resulting in significant re-exposure.

\subsection{Comparison of information diffusion strategies}
\begin{figure}
\centering
\includegraphics[width=16cm]{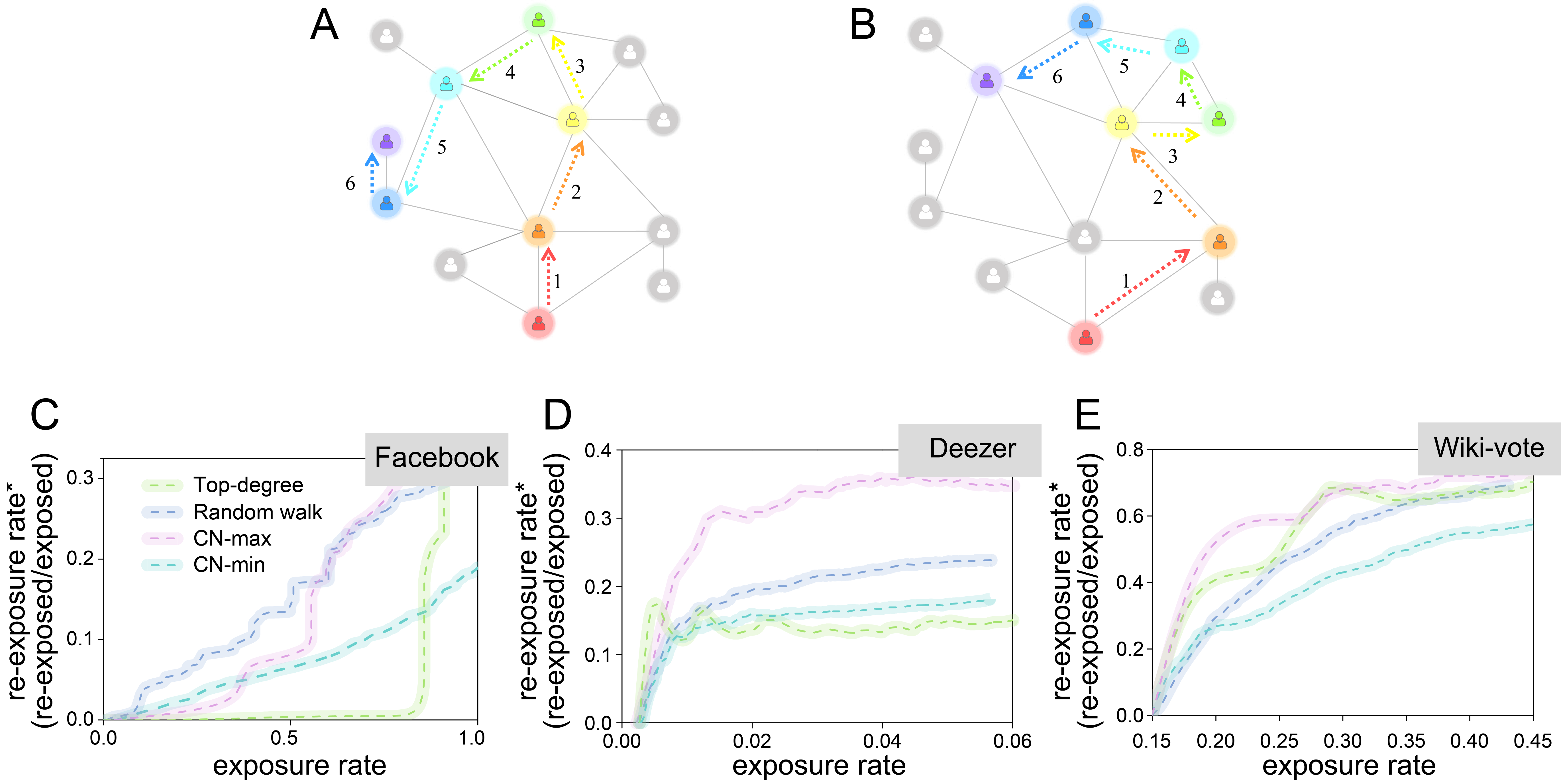}
\caption{Information diffusion based on common neighbor strategies. ($A$) The CN-max strategy spreads information along links with the largest number of common neighbors. ($B$) The CN-min strategy disseminate information along links with the minimum number of common neighbors. ($C$) to ($E$) shows exposure and re-exposure rate of information spreading under different propagation strategies.}
\label{fig:4}
\end{figure}

The above simulation results and empirical observations thus indicate that the effectiveness of information propagation can be substantially different through different patterns of transmission, as well as along users with unique local personal network structures. We compare the proposed CN strategy (see Materials and methods) with random walk (RW), in which the information starts from a seed node and at each step propagates randomly to a friend of the current publisher, and targeted ``delivery'' strategies, in which a set of nodes are selected according to specific criteria, e.g., top degree (TD), to investigate the performance of different policies on the exposure and re-exposure rate in information spreading (Fig.~\ref{fig:4}$C$-$E$). In all empirical networks, the CN-max policy can produce the highest $R^{\rm exp}$ for the given exposure rate. Particularity, in the Deezer network, $R^{\rm exp}$ of CN-max reaches 15\% higher than the next best strategy (RW). On the other hand, the CN-min strategy produces much lower $R^{\rm exp}$ for any given $R^{\rm re}$. In addition, it is worth noting that the global strategy of TD has a very different performance in the three networks: for a broad range of $R^{\rm exp}$ < 45\%, the re-exposure rate $R^{\rm re}$ can be restricted close to zero, which means that when using the most connected nodes to disseminated information without propagation, almost no one can see the information repeatedly even when half of the nodes have this information visible in their social networks; the TD strategy has low performance in the Deezer network, however, in the Wiki-vote network, the performance of TD is vary similar to CN-max, that for any given exposure rate, it produces almost the highest $R^{\rm re}$. 

Furthermore, the different performance of TD strategy on three networks implies that this strategy strongly depends on the structure of networks, in contrast to the relatively robust performance of the RW, CN-max, and CN-min strategies. It is clear that the community structure of the Facebook network is the main reason for TD to achieve a high $R^{\rm exp}$ with low $R^{\rm re}$, as the information posted by the most connected nodes are scattered in weakly connected communities, hardly to form mutual visibility for producing redundancy (Fig.~S3).

\section{Discussion}
In this study, we have extended the traditional research of information spreading models by considering the coverage (exposure) and multiple influences (re-exposure) simultaneously. We show that the network structure, as well as the transmission pathways, are critical for the two metrics to be optimized. Generally, users with more connections in the network tend to be associated with a higher probability of getting overwhelmed with popular information, and that clustering and community structure prohibits long transmission waves while keeping social reinforcement effect strong. We have also identified a few representative strategies which may be implemented to minimize or maximize repeated exposure. We believe that many of the behavioral patterns we see emerge on social networks, both positive and negative ones are driven by the fact that exposure and re-exposure rate demonstrate different characteristics. In fact, one may use the different re-exposure rates to identify and better segment types of information and kind of influencers. 

The rapid development of online social media has brought tremendous changes to business operations. The widespread popularity of online social media has enabled marketers to gradually reduce the dependency on traditional media outlets and instead use word-of-mouth as a method to motivate users to disseminate information. On the other hand, controlling information overload and duplication are also goals for platforms to improve user experience and maintain platform activity. Although these contradictory goals can be optimized under the principles and strategies proposed in this study, it is, however, essential to keep in mind that the cost should also be taken into account for further implementations.

\section{Materials and methods}
\subsection{Network data}
We use four empirical online social networks for the investigation of the impact of structural parameters and pattern of transmission on the exposure and re-exposure rate of information spreading, including the dataset of 
\href{http://snap.stanford.edu/data/}{Facebook}, \href{http://snap.stanford.edu/data/}{Deezer}, \href{http://snap.stanford.edu/data/}{Wikipedia-Vote} and \href{http://www.weibo.com}{Sina Weibo}. The statistics of these networks are listed in Table S1 and descriptions on these datasets are presented in {{\textit {SI Appendix}}, section \uppercase\expandafter{\romannumeral3}}.

\subsection{Approximation of the individual exposure and re-exposure probability}
Consider a random walk on $G$, the visitation probability of each node $P_i$, can be derived by the equilibrium of a Markov process with a transition matrix based on the adjacency and degree of nodes in the network \cite{lu2012sensitivity}. It can be verified that, when the network is undirected and connected, the probability of forwarding the information at each node is asymptotically proportional to its degree \cite{volz2008probability}, i.e., $P_i=d_i/\sum{^N_{j=1}}d_j$, where $d_i$ is the degree of $v_i$. After $n$ steps, we can approximate the probability $Pr_i$ that $v_i$ becomes a re-exposed node is:
\begin{equation}
Pr_i=Pc_i-C_n^1 \mu_i \alpha_i^{n-1}-\delta _i
\end{equation}
where $Pc_i$ is the probability that $v_i$ is an exposed node, $\mu_i$ and $\alpha_i$ are the probability that exactly one neighbor of $v_i$ posts information and any neighbor of $v_i$ do not post information at the initial step, respectively(see {{\textit {SI Appendix}}, section \uppercase\expandafter{\romannumeral2}}). 
$\delta_i$ is the part of being repeatedly calculated when only one neighbor around $v_i$ posts information:
\begin{equation}
\delta _i=\sum_{l=1}^{d_i}{ \sum_{t=2}^{n} [ C_n^t {P_{k_l}}^t( 1-P_{k_l})^{n-t}\prod_{m=1,m\neq l}^{d_i}( 1-P_{k_l} )^n]}
\end{equation}

\subsection{Information diffusion strategy}
More redundant ties better promote the diffusion of behaviors across large population \cite{centola2007cascade}. A competing hypothesis argues that when behaviors require social reinforcement, a network with more clustering may be more advantageous. This hypothesis predicts that clustered-lattice networks have more redundant ties. Therefore, we develop an advertising delivery strategy, which uses the common neighbors (CN) indicator \cite{lorrain1971structural} to design the path of information dissemination (Fig.~\ref{fig:4}$A$-$B$). The CN indicator is defined as follows: For a node $v_x$ in the network, defining its neighbor set as $\Gamma(x)$, the similarity of the two nodes $v_x$ and $v_y$ is defined as their number of common neighbors:
\begin{equation}
CN_{xy}=\left | \Gamma (x)\cap \Gamma (y) \right |
\end{equation}
The number of common neighbors is equal to the number of paths between the two nodes with length two, i.e., $CN_{xy} = (A^2)_{xy}$. In the spreading process, we then allow the information either traverses along the links with the minimum (CN-Min) or the maximum (CN-Max) number of common neighbors, to achieve different redundancy rate while keeping the coverage fixed.

\textbf{Acknowledgements}
XL acknowledges the Natural Science Foundation of China under Grant 71771213, 71522014 and 71731009. SQ was partially supported by the Natural Science Foundation of China under Grant 71790615, and 71725001. PH was supported by JSPS KAKENHI (JP 18H01655). The funders had no part in the design of this research and the opinions expressed here do not necessarily reflect the views of the funders.

\bibliographystyle{unsrt}  
\bibliography{references.bib}

\begin{thebibliography}{10}

\bibitem{morone2015influence}
Flaviano Morone and Hern{\'a}n~A Makse.
\newblock Influence maximization in complex networks through optimal
  percolation.
\newblock {\em Nature}, 524(7563):65, 2015.

\bibitem{aral2012identifying}
Sinan Aral and Dylan Walker.
\newblock Identifying influential and susceptible members of social networks.
\newblock {\em Science}, 337(6092):337--341, 2012.

\bibitem{kitsak2010identification}
Maksim Kitsak, Lazaros~K Gallos, Shlomo Havlin, Fredrik Liljeros, Lev Muchnik,
  H~Eugene Stanley, and Hern{\'a}n~A Makse.
\newblock Identification of influential spreaders in complex networks.
\newblock {\em Nature Physics}, 6(11):888, 2010.

\bibitem{hu2018local}
Yanqing Hu, Shenggong Ji, Yuliang Jin, Ling Feng, H~Eugene Stanley, and Shlomo
  Havlin.
\newblock Local structure can identify and quantify influential global
  spreaders in large scale social networks.
\newblock {\em Proceedings of the National Academy of Sciences},
  115(29):7468--7472, 2018.

\bibitem{censor2010partial}
Keren Censor~Hillel and Hadas Shachnai.
\newblock Partial information spreading with application to distributed maximum
  coverage.
\newblock In {\em Proceedings of the 29th ACM SIGACT-SIGOPS symposium on
  Principles of Distributed Computing}, pages 161--170. ACM, 2010.

\bibitem{edmunds2000problem}
Angela Edmunds and Anne Morris.
\newblock The problem of information overload in business organisations: a
  review of the literature.
\newblock {\em International Journal of Information Management}, 20(1):17--28,
  2000.

\bibitem{iribarren2011branching}
Jos{\'e}~Luis Iribarren and Esteban Moro.
\newblock Branching dynamics of viral information spreading.
\newblock {\em Physical Review E}, 84(4):046116, 2011.

\bibitem{hinz2011seeding}
Oliver Hinz, Bernd Skiera, Christian Barrot, and Jan~U Becker.
\newblock Seeding strategies for viral marketing: An empirical comparison.
\newblock {\em Journal of Marketing}, 75(6):55--71, 2011.

\bibitem{adamopoulos2015effectiveness}
Panagiotis Adamopoulos and Vilma Todri.
\newblock The effectiveness of marketing strategies in social media: Evidence
  from promotional events.
\newblock In {\em Proceedings of the 21th ACM SIGKDD International Conference
  on Knowledge Discovery and Data Mining}, pages 1641--1650. ACM, 2015.

\bibitem{bapna2015your}
Ravi Bapna and Akhmed Umyarov.
\newblock Do your online friends make you pay? a randomized field experiment on
  peer influence in online social networks.
\newblock {\em Management Science}, 61(8):1902--1920, 2015.

\bibitem{nielsen2012consumer}
Nielsen.
\newblock Consumer trust in online, social and mobile advertising grows.
\newblock
  \url{https://www.nielsen.com/us/en/insights/article/2012/consumer-trust-in-online-social-and-mobile-advertising-grows/},
  2012.

\bibitem{local2014local}
Bright Local.
\newblock Local consumer review survey.
\newblock
  \url{https://www.brightlocal.com/research/local-consumer-review-survey/},
  2018.

\bibitem{bakshy2012the}
Eytan {Bakshy}, Itamar {Rosenn}, Cameron {Marlow}, and Lada~A. {Adamic}.
\newblock The role of social networks in information diffusion.
\newblock In {\em Proceedings of the 21st International Conference on World
  Wide Web}, pages 519--528, 2012.

\bibitem{centola2010the}
Damon {Centola}.
\newblock The spread of behavior in an online social network experiment.
\newblock {\em Science}, 329(5996):1194--1197, 2010.

\bibitem{zheng2013spreading}
Muhua Zheng, Linyuan L{\"u}, Ming Zhao, et~al.
\newblock Spreading in online social networks: The role of social
  reinforcement.
\newblock {\em Physical Review E}, 88(1):012818, 2013.

\bibitem{granovetter1978threshold}
Mark {Granovetter}.
\newblock Threshold models of collective behavior.
\newblock {\em American Journal of Sociology}, 83(6):1420--1443, 1978.

\bibitem{Romero2011}
Daniel~M. Romero, Brendan Meeder, and Jon Kleinberg.
\newblock Differences in the mechanics of information diffusion across topics:
  Idioms, political hashtags, and complex contagion on twitter.
\newblock In {\em Proceedings of the 20th International Conference on World
  Wide Web}, pages 695--704. ACM, 2011.

\bibitem{zhang2016dynamics}
Zi-Ke {Zhang}, Chuang {Liu}, Xiu-Xiu {Zhan}, Xin {Lu}, Chu-Xu {Zhang}, and
  Yi-Cheng {Zhang}.
\newblock Dynamics of information diffusion and its applications on complex
  networks.
\newblock {\em Physics Reports}, 651:1--34, 2016.

\bibitem{pearson1905the}
Karl {Pearson}.
\newblock The problem of the random walk.
\newblock {\em Nature}, 72(1865):294--294, 1905.

\bibitem{ney1965the}
P.~E. {Ney}.
\newblock The theory of branching processes.
\newblock {\em Technometrics}, 7(1):79--80, 1965.

\bibitem{rogers1995diffusion}
Everett~M Rogers.
\newblock Diffusion of innovations: modifications of a model for
  telecommunications.
\newblock In {\em Die diffusion von innovationen in der telekommunikation},
  pages 25--38. Springer, 1995.

\bibitem{pennings1973measures}
Johannes {Pennings}.
\newblock Measures of organizational structure: A methodological note.
\newblock {\em American Journal of Sociology}, 79(3):686--704, 1973.

\bibitem{watts2000small}
Duncan~J. {Watts} and Per {Bak}.
\newblock Small worlds: The dynamics of networks between order and randomness.
\newblock {\em Physics Today}, 53(11):54--55, 2000.

\bibitem{granovetter1973the}
Mark~S. {Granovetter}.
\newblock The strength of weak ties.
\newblock {\em American Journal of Sociology}, 78(6):1360--1380, 1973.

\bibitem{leskovec2007dynamics}
Jure Leskovec, Lada~A Adamic, and Bernardo~A Huberman.
\newblock The dynamics of viral marketing.
\newblock {\em ACM Transactions on the Web (TWEB)}, 1(1):5, 2007.

\bibitem{liben2008tracing}
David Liben-Nowell and Jon Kleinberg.
\newblock Tracing information flow on a global scale using internet
  chain-letter data.
\newblock {\em Proceedings of the National Academy of Sciences},
  105(12):4633--4638, 2008.

\bibitem{lu2012sensitivity}
Xin Lu, Linus Bengtsson, Tom Britton, Martin Camitz, Beom~Jun Kim, Anna
  Thorson, and Fredrik Liljeros.
\newblock The sensitivity of respondent-driven sampling.
\newblock {\em Journal of the Royal Statistical Society: Series A (Statistics
  in Society)}, 175(1):191--216, 2012.

\bibitem{volz2008probability}
Erik Volz and Douglas~D Heckathorn.
\newblock Probability based estimation theory for respondent driven sampling.
\newblock {\em Journal of Official Statistics}, 24(1):79, 2008.

\bibitem{centola2007cascade}
Damon {Centola}, Víctor~M. {Eguíluz}, and Michael~W. {Macy}.
\newblock Cascade dynamics of complex propagation.
\newblock {\em Physica A-statistical Mechanics and Its Applications},
  374(1):449--456, 2007.

\bibitem{lorrain1971structural}
François {Lorrain} and Harrison~C. {White}.
\newblock Structural equivalence of individuals in social networks.
\newblock {\em Journal of Mathematical Sociology}, 1(1):49--80, 1971.

\bibitem{kermack1927contribution}
William~Ogilvy Kermack and Anderson~G McKendrick.
\newblock A contribution to the mathematical theory of epidemics.
\newblock {\em Proceedings of the royal society of london. Series A, Containing
  papers of a mathematical and physical character}, 115(772):700--721, 1927.

\bibitem{kermack1932contributions}
William~Ogilvy Kermack and Anderson~G McKendrick.
\newblock Contributions to the mathematical theory of epidemics. ii.—the
  problem of endemicity.
\newblock {\em Proceedings of the Royal Society of London. Series A, containing
  papers of a mathematical and physical character}, 138(834):55--83, 1932.

\bibitem{hastings1970monte}
W~Keith Hastings.
\newblock Monte carlo sampling methods using markov chains and their
  applications.
\newblock {\em Biometrika}, 57(1):97--109, 1970.

\bibitem{heckathorn20076}
Douglas~D Heckathorn.
\newblock 6. extensions of respondent-driven sampling: Analyzing continuous
  variables and controlling for differential recruitment.
\newblock {\em Sociological methodology}, 37(1):151--208, 2007.

\bibitem{leskovec2012learning}
Jure Leskovec and Julian~J Mcauley.
\newblock Learning to discover social circles in ego networks.
\newblock In {\em Advances in neural information processing systems}, pages
  539--547, 2012.

\bibitem{leskovec2010predicting}
Jure Leskovec, Daniel Huttenlocher, and Jon Kleinberg.
\newblock Predicting positive and negative links in online social networks.
\newblock In {\em Proceedings of the 19th international conference on World
  wide web}, pages 641--650. ACM, 2010.

\bibitem{rozemberczki2018gemsec}
Benedek Rozemberczki, Ryan Davies, Rik Sarkar, and Charles Sutton.
\newblock Gemsec: graph embedding with self clustering.
\newblock {\em arXiv preprint arXiv:1802.03997}, 2018.

\bibitem{hjouji2018impact}
Zakaria~el Hjouji, D~Scott Hunter, Nicolas Guenon~des Mesnards, and Tauhid
  Zaman.
\newblock The impact of bots on opinions in social networks.
\newblock {\em arXiv preprint arXiv:1810.12398}, 2018.

\bibitem{parlapiano2018propaganda}
Alicia Parlapiano and Jasmine~C Lee.
\newblock The propaganda tools used by russians to influence the 2016 election.
\newblock {\em The New York Times}, 2018.

\bibitem{shao2018spread}
Chengcheng Shao, Giovanni~Luca Ciampaglia, Onur Varol, Kai-Cheng Yang,
  Alessandro Flammini, and Filippo Menczer.
\newblock The spread of low-credibility content by social bots.
\newblock {\em Nature communications}, 9(1):4787, 2018.

\end{thebibliography}

\newpage
\textbf{\huge{Supplementary Information}}
\subsection*{\uppercase\expandafter{\romannumeral1.} Random walk and branching process}
In a random walk, the information spreads from a seed node.  In each subsequent step, it is re-tweeted by a randomly selected friends of the ancestor node. In addition, we consider the scenario that at each wave of transmission, the information can be forwarded by more than one friend. We herein adopt the widely used SI model, in which each node in the network can be in either the S state or the I state \cite{kermack1927contribution,kermack1932contributions}. When a node has seen the information shared by its friends, it will, with probability $\beta$ adopt and re-tweet this information. To approximate the reality that the information shared by each users' friends updates quickly, such that one will not consider whether or not to post an information back and forth, we assume that each node will make only one decision on re-tweeting when the information appears in its social network at the first time it observes the information, and the state of the node will not be affected and changed in subsequent propagation.  

\subsection*{\uppercase\expandafter{\romannumeral2.} Approximation of the exposure and re-exposure probability}
Consider a random walk on $G$, the visitation probability of each node $P_i$, can be derived by the equilibrium of a Markov process with transition matrix $B$ \cite{hastings1970monte,heckathorn20076, lu2012sensitivity, volz2008probability}:

\begin{equation}
P^TB=P^T
\end{equation}
in which:

\begin{equation}
B=
\begin{pmatrix}
0 & a_{12}/d_1^o & \cdots  & a_{1N}/d_1^o\\ 
a_{21}/d_2^o & 0 & \cdots  & a_{2N}/d_2^o\\
\vdots  & \vdots  & \ddots   & \vdots \\ 
a_{N1}/d_N^o  &a_{N2}/d_N^o & \cdots  & 0
\end{pmatrix}
\end{equation}
where $d_i^o$ is the out-degree of $v_i$. Eq. 2 indicates that $P$ is the eigenvector for eigenvalue 1 for $B^T$. It can be verified that, when the network is undrected and connected, $P_i=d_i/\sum{^N_{j=1}}d_j$ is the sulotion of Eq. 1, i.e., under random walk model, the probability of forwarding the information at each node is asymptotically proportional to its degree. When the number of branches at each step is small, the branching process can also be approximated by the model presented here, as illustrated in \cite{lu2012sensitivity,volz2008probability}.

With the visitation probability $P_i$, denote $K=\left \{ K_1,K_2,\cdots,k_{d_i} \right \}$ be the serial number of the current neighbors of $v_i$. At step $n=1$, the probability $c_i$ that node $v_i$ is exposed can be written as:

\begin{equation}
c_i=1-\left ( 1-p_i \right )\left [ \prod_{l=1}^{d_j} \left ( 1-P_{k_l} \right )\right]
\end{equation}
where $\alpha_i=\prod_{l=1}^{d_j} \left ( 1-P_{k_l} \right )$ is the probability that any neighbor of $v_i$ does not share the information. After $n$ steps, the probability $Pc_i$ that $v_i$ is an exposed node can be approximated by:

\begin{equation}
Pc_i=1-\left (1-c_i \right )^n-\sum_{t=4}^{n}2\left ( n-1 \right )n^{t-3}+\left ( n-1 \right )^2n^{t-4}(t-3)
\end{equation}
similarly, at step $n=1$, the probability $r_i$ that $v_i$ is a re-exposed node can be calculated by:

\begin{equation}
r_i=1-c_i-\sum_{l=1}^{d_i}\left [ P_{k_l}\prod_{m=1\\m\neq l}^{d_i}\left ( 1-P_{k_l} \right ) \right ]
\end{equation}
where $\mu_i=\sum_{l=1}^{d_i}\left [ P_{k_l}\prod_{m=1,m\neq l}^{d_i}\left ( 1-P_{k_l} \right ) \right ]$ is the probability that exactly one neighbor of $v_i$ posts the information. After $n$ steps, the probability $Pr_i$ that $v_i$ becomes a re-exposed node is:
\begin{equation}
Pr_i=Pc_i-C_n^1 \mu_i \alpha_i^{n-1}-\delta _i
\end{equation}
where $\delta_i$ accounts for the repeatedly calculated part when only one neighbor around $v_i$ posts the information:

\begin{equation}
\delta _i=\sum_{l=1}^{d_i}{ \sum_{t=2}^{n} [ C_n^t {P_{k_l}}^t( 1-P_{k_l})^{n-t}\prod_{m=1,m\neq l}^{d_i}( 1-P_{k_l} )^n]}
\end{equation}

\subsection*{\uppercase\expandafter{\romannumeral3.} Data collection}
The Facebook dataset was collected from survey of participants that use Facebook. The data contains for each user, the ``circles'' or ``friend list'', which is then used to construct a social network with 4,039 nodes and 88,234 links \cite{leskovec2012learning}.

Deezer is an Internet-based music streaming service which was created in Paris and it has 14 million monthly active users. This data was collected in November 2017, containing the friendship networks of users from a European country. The network includes 41,773 nodes and 125,826 links \cite{leskovec2010predicting}.

In order for a user to become an administrator, the Wikipedia community via a public discussion or a vote decides who to promote. Wiki-vote contains all the voting history data from the Wikipedia till January 2008. A directed edge from node $i$ to node $j$ represents that user $i$ voted on user $j$. The network includes 7,115 nodes and 103,689 links \cite{rozemberczki2018gemsec}.

Sina Weibo is the most popular twitter-like social media platform in China. By using a fast web-crawling framework and elaborately designing the crawler, we obtained the network structure of 430 million users until 2017, including the number of users' followings and followers. The network includes 430 million (432,908,953) nodes and 51.8 billion (51,862,490,000) links. We crawled a randomly set of selected popular news posts of Sina Weibo platform in the second half of 2017, and the number of forwardings ranges from 10,000 to 100,000. According to the data set, we are able to construct the trajectory data for news re-posting/re-tweeting based on the current user's re-post time and the re-post identifier ``@''.

\subsection*{\uppercase\expandafter{\romannumeral4.} Simulation setting}
To evaluate the validity of $Pr_i$, we implement both a simulated random walk and branching processes on BA networks with size $N=10,000$ and average degree $<d>=4$. In each simulation, the seed node is randomly chosen. The spreading is continued until half of the nodes in the network have re-posted, i.e., become state I.

Before the examination of exposure and re-exposure rate of real-life information spreading on Sina Weibo, we implement the aforementioned branching process to examine the effect of network structure parameters. On Facebook, Deezer, and Wiki-vote, simulated information is disseminated until about 20\% of the nodes in the network have re-tweeted. Proper transmission probability $\beta \subseteq \left ( 0,0.3 \right )$is selected to guarantee the effectiveness of the spreading, i.e., the information won't die out before reaching the desired outbreak scale. For each network, the simulations are repeated 100 times.

\subsection*{\uppercase\expandafter{\romannumeral5.} Correlation between nodal degree and information redundancy on Sina Weibo}
There is a surprisingly drop of $R^{\rm exp}$ and $R^{\rm re}$ for nodes with degrees in the region of ($2^{11}$, $2^{13}$] when we analyze the correlation between network structure and  $R^{\rm exp}$, $R^{\rm re}$, indicating that nodes within this region have substantially different local networks in comparison to ``normal'' users. This phenomenon can be attributed to a variety of reason, including the limit of 2,000 (set by the platform) for the number of people each ordinary user can follow, and it may also be caused by automated accounts or bots who post or repost message repeatedly. 

First, on Sina Weibo, there is a limit of 2,000 for the number of people each ordinary user can follow. If a user needs to follow more than 2,000 people, an official application and certification process is required by the platform. As Sina Weibo has become one of the most active online social media, it has also been used as a popular marketing channel in recent years. Companies and self- employed marketers are increasingly engaged in Sina Weibo for advertising and product promotion. In the hope of being following back, achieving good publicity and promotion effects, their accounts usually follow a large number of unknown users. However, due to the lack of social networking property, i.e., most of followers who belong to users in the region of ($2^{11}$, $2^{13}$] do not know each other, the pathways of information spreading can hardly pass along the social circles of such accounts, leading to extremely low exposure rate and re-exposure rate.

Secondly, we find that both $R^{\rm exp}$ and $R^{\rm re}$  are the lowest in the region of ($2^{11}$, $2^{13}$] while the nodal degree shows two peaks at $out-degree=4,143$ and $out-degree=4,607$. This phenomenon is attributed to automated accounts or bots who post or re-post message repeatedly. A deeper investigation also reveals that their active time concentrated in the early mornings. This phenomenon coincides with the latest research, such as Zakaria et al.\ which presents an analysis of the impact of automated accounts, or bots, on opinions in a social network. They find a small number of highly active bots in a social network can have a disproportionate impact on opinions \cite{hjouji2018impact}. It is suspected that the 2016 U.S. presidential election was the victim of such social network interference, potentially by foreign actors \cite{parlapiano2018propaganda}. Social bots played a major role in spreading articles from low-credibility sources \cite{shao2018spread}. All of these findings indicate that the increasingly ``anthropomorphic'' bots in social networks have gradually become a part of social users. From the managerial perspectives, this finding can provide some constructive comments on the control of redundant information on online social platforms such as strengthen the control of the automated accounts and bots. Owing to the fact that there is a limit of the number of influential users (such as pop stars who was verified by Twitter, influential, and with numerous followers). While the number of users' followings reaches 10,000 or more, this kind of users are easily covered by popular information and to be exposed or re-exposed.

\newpage
\begin{table}
\renewcommand\thetable{S1}
\centering
\caption{Basic statistics of the four empirical social networks, which have $N$ nodes and $M$ edges. $<d>$ is the average degree. $d_{max}$ is the maximum degree. $<c>$ represents the average clustering coefficient.}
\begin{tabular}{cccccc}
Network & $N$ & $M$ & $\langle d\rangle$ & $d_{max}$ & $<c>$ \\
\midrule
Facebook & 4,039 & 88,234 & 43 & 1,045 & 0.6055 \\
Deezer & 41,773 & 125,826 & 6 & 112 & 0.0912 \\
Wiki-Vote & 7,115 & 103,689 & 28 & 1,065 & 0.081\\
Sina Weibo & 432,908,953 & 51,862,490,000 & 120 & 19,860 & 0 \\
\bottomrule
\end{tabular}
\end{table}

\newpage
\begin{figure}
\renewcommand\thefigure{S1}
\centering
\includegraphics[width=\textwidth]{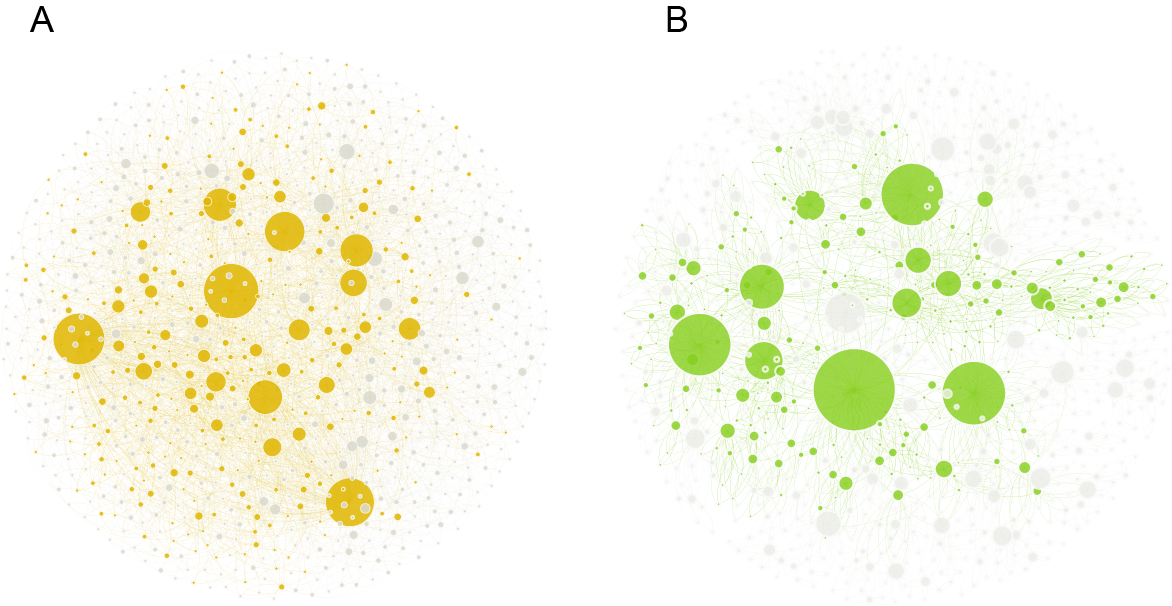}
\caption{Visualization of information diffusion in scale-free networks with different global clustering coefficient. ($A$) $C$=0.184; ($B$) $C$=0.718. The networks are configured with $N$ = 1,000 and $D$ = 4; the size of spreading is about 300.}
\label{fig:S1}
\end{figure}

\newpage
\begin{figure}
\renewcommand\thefigure{S2}
\centering
\includegraphics[width=\textwidth]{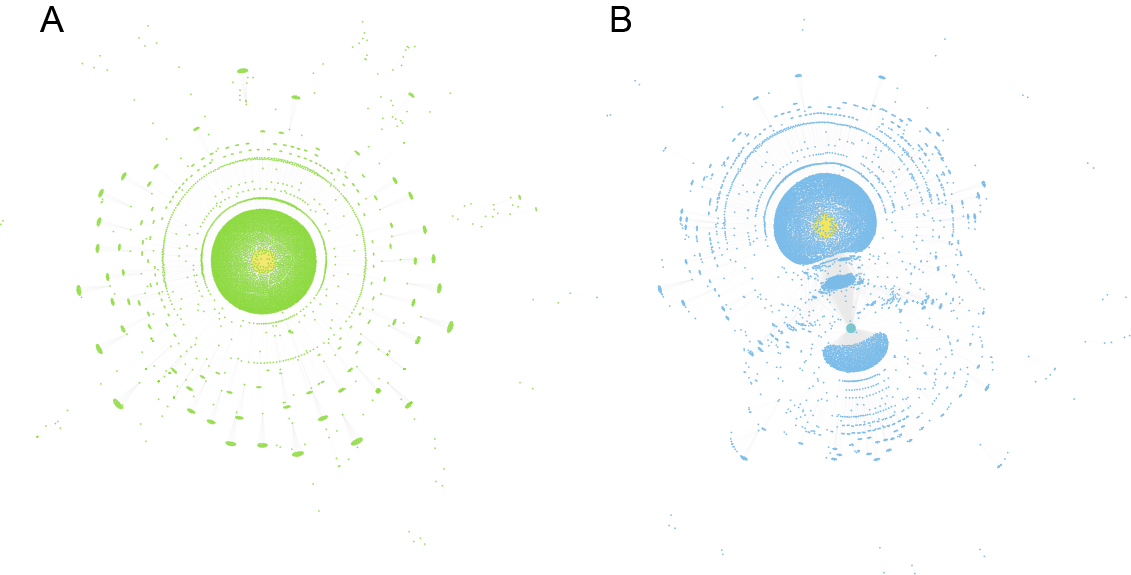}
\caption{Illustration on the distinct transmission pathways of type I and type II news on Sina Weibo. ($A$) A type I news which was published initially by the famous Chinese singer Gloria Tang (G.E.M. the large node in the middle), when she participated in the show ``I am singer''; ($B$) A type II news which was published by a less influential node (the lower large node) and get forwarded by an influential user Aaron Yan.}
\label{fig:S2}
\end{figure}

\newpage
\begin{figure}
\renewcommand\thefigure{S3}
\centering
\includegraphics[width=\textwidth]{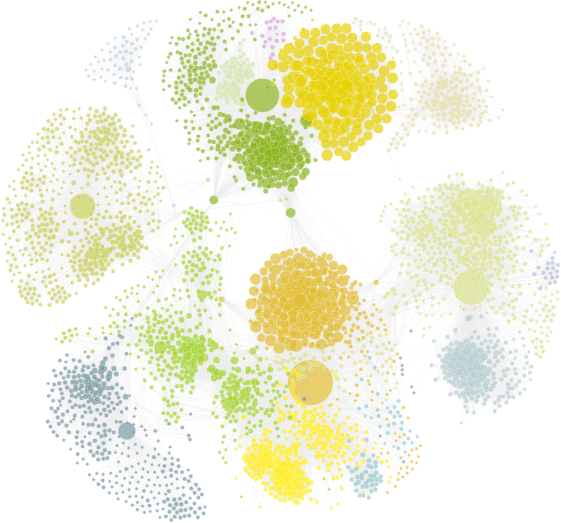}
\caption{Visualization of the Facebook network.}
\label{fig:S3}
\end{figure}

\end{document}